\documentclass[]{elsart}

\usepackage{dcolumn}
\usepackage{graphicx}
\usepackage{amsmath}
\usepackage{amssymb}
\usepackage{bm}

\newcommand{\be}{\begin{equation}}
\newcommand{\ee}{\end{equation}}
\newcommand{\bea}{\begin{eqnarray}}
\newcommand{\eea}{\end{eqnarray}}

\begin{document}

\bibliographystyle{myprsty}

\begin{frontmatter}

\title{Reparameterization invariance of NRQED\\
self-energy corrections and improved theory\\ 
for excited D states in hydrogenlike systems}

\newcommand{\addrHDUNI}{Institute f\"ur Theoretische Physik,
Philosophenweg 16, 69120 Heidelberg, Germany}

\newcommand{\addrHDMPI}{Max--Planck--Institut f\"ur
Kernphysik, Postfach 103980, 69029 Heidelberg, Germany}

\author{Benedikt J. Wundt}
\address{Max--Planck--Institut f\"ur
Kernphysik, Postfach 103980, 69029 Heidelberg, Germany}
\author{Ulrich D. Jentschura}
\address{Max--Planck--Institut f\"ur
Kernphysik, Postfach 103980, 69029 Heidelberg and \\
Institut f\"ur Theoretische Physik,
Philosophenweg 16, 69120 Heidelberg}

\date{\today}

\begin{abstract}
Canonically, the quantum electrodynamic radiative 
corrections in bound systems have been evaluated 
in photon energy regularization, i.e.~using a noncovariant 
overlapping parameter that separates the high-energy 
relativistic scales of the virtual quanta from the nonrelativistic 
domain. Here, we calculate the higher-order corrections to the one-photon 
self-energy calculation with three different overlapping 
parameters (photon energy, photon mass and 
dimensional regularization) and demonstrate the reparameterization
invariance of nonrelativistic quantum electrodynamics (NRQED) using
this particular example. We also present new techniques for the calculation
of the low-energy part of this correction, which lead to 
results for the Lamb shift of highly excited states that 
are important for high-precision spectroscopy.
\end{abstract}


\end{frontmatter}

\newpage

{\it {Introduction.---}}
In 1986, Caswell and Lepage introduced the idea of nonrelativistic quantum
electrodynamics (NRQED) to describe bound states~\cite{CaLe1986}. 
The underlying notion is to reformulate the predictions of full 
relativistic quantum electrodynamics in terms of operators 
acting on nonrelativistic spinors, so that the higher-order corrections
can be expressed in terms of a series of iterated operators of 
lower order, and additional operators which come in at every given 
order in the so-called $Z\alpha$ expansion, where $Z$ is the nuclear
charge number, and $\alpha$ is the fine-structure constant.

The development of NRQED seems to have been 
motivated by the well-known fact that a ``usual'' quantum field theory 
based on $S$-matrix elements evaluated on free states and free 
propagators is not enough to describe bound states. 
Two energy scales are present in the problem, and these 
have to be separated by a so-called overlapping parameter
(see \S~123 of Ref.~\cite{BeLiPi1982}). The two energy domains
are the high-energy relativistic scale of the virtual 
quanta and the nonrelativistic
domain of bound-state momenta and energies. 

Note that the regularization and renormalization
of amplitudes in the ultraviolet (UV) at
some mass scale $\Lambda_{\rm UV}$ has got nothing to do with the 
scale-separation, or overlapping, parameter.
For the overlapping parameter, one can 
has use either a photon energy $\epsilon$, or a photon mass $\mu$,
or one can work in $4-2 \varepsilon_D$ space-time dimensions
(we emphasize that $\epsilon \neq \varepsilon_D$, the index 
$D$ is supplied in this work in order to facilitate the distinction 
of the two regularizations).
Feynman~\cite{Fe1949}, as well as 
French and Weisskopf~\cite{FrWe1949} did their 
calculations in photon energy regularization.
It turned out to be nontrivial to convert the high-energy part 
from a photon mass $\mu$ to a photon energy $\epsilon$
(see the rather well-known footnote 13 on p.~777 of Ref.~\cite{Fe1949}),
while the photon energy regularization 
is the most natural cutoff for the low-energy part. 

All three regularizations have been used in the literature
for the treatment of different bound-state QED problems.
E.g., Nio and Kinoshita~\cite{NiKi1997} used
photon mass regularization for their calculation of the 
higher-order binding corrections to the muonium hyperfine 
structure, while Pachucki~\cite{Pa1996} 
used photon energy regularization for the same problem.
Dimensional regularization has been described for the 
lowest-order Lamb shift 
by Pineda and Soto in 1998~\cite{PiSo1998}. Higher-order 
binding corrections to the Lamb shift have been evaluated 
in dimensional regularization in~\cite{JeCzPa2005}.

This Letter has a twofold purpose. 
(i)~Here, for the first time to the best of our knowledge, a calculation
of a nontrivial QED correction is presented in 
all three common (re-)parameterizations of NRQED (photon energy,
photon mass and dimensional regularization). Namely, 
we consider the higher-order 
binding corrections to the one-loop self-energy in hydrogenlike systems.
We thereby verify the reparameterization invariance to the full extent,
for all three common regularization methods, while working 
on the same problem employing three different methods.
(ii)~As a byproduct, we develop methods to do the calculation of 
relativistic Bethe logarithms for arbitrary Rydberg states
of hydrogenlike systems, and we evaluate these 
corrections for states with principal quantum numbers as high 
as $n=12$, where the excessive number of terms has been 
prohibiting both numerical as well as analytic approaches
in the past. 

{\em {Reparameterization invariance (general remarks).---}} 
Our final goal is to express the self-energy shift $\Delta E(nL_j)$ of a 
general hydrogenic state with orbital angular momentum quantum 
number $L \geq 2$, total angular momentum $j$ and 
principal quantum number $n$, 
\begin{equation}
\label{DELTAE}
\Delta E(nL_j) = \frac{\alpha}{\pi}\, \frac{(Z\alpha)^4}{n^3} \, 
F(nL_j) \,,
\end{equation}
in terms of the reference state quantum numbers
(we use natural units with $\hbar = c = \epsilon_0 = 1$, and 
we choose the energy scale so that the electron mass $m = 1$). The scaled 
self-energy function $F(nL_j)$ has the following 
semi-analytic expansion (it is not analytic because of the presence
of logarithms),
\begin{equation}
\label{ACES}
F(nL_j) = A_{40} + (Z\alpha)^2 \, \left\{ A_{61} \, 
\ln[(Z\alpha)^{-2}] + A_{60} \right\}\,,
\end{equation}
where the first index of the $A$ coefficients counts the number
of factors $Z\alpha$, whereas the second counts the power of 
the logarithm $\ln[(Z\alpha)^{-2}]$.

The reparameterization invariance of NRQED implies that 
the $F$ function should be expressible as the sum
of a regularized high-energy part $F_H$ 
and a regularized low-energy part $F_L$, where $F_H$ and $F_L$
can be formulated in photon energy, photon mass
or in dimensional regularization, 
as follows,
\begin{equation}
F = F_H(\epsilon) + F_L(\epsilon) 
= F_H(\mu) + F_L(\mu) 
= F_H(\varepsilon_D) + F_L(\varepsilon_D) \,.
\end{equation}

{\em {High-energy part.---}}In the treatment of the 
one-loop self-energy, we start with the high-energy part, which 
corresponds to photon energies of the order of the electron mass, and 
electron momenta of the order of the atomic momenta $Z\alpha$,
where $Z$ is the nuclear charge, and $\alpha$ is the fine-structure 
constant. We identify all operators that contribute 
at the order $\alpha (Z\alpha)^6$, and evaluate these for a 
general state in a hydrogenlike system with orbital angular momentum
quantum number $L \geq 2$, in photon energy, photon mass, and also 
in dimensional regularization. We find that the final expressions
simplifies considerably for these states, 
and indeed very compact final results can be indicated.

The different contributions to the high-energy part, for 
states with novanishing orbital angular momentum, can be described as 
follows, in terms of the electron Dirac form factor 
$F_1$ and the electron magnetic form factor $F_2$. 
Here, we give only an indication of these corrections, 
detailed formulas corresponding to the terms mentioned below can 
be found in Ref.~\cite{JeCzPa2005}.
First, we have an $F'_1(0)$ correction evaluated on the 
relativistic wave function, where the latter has to be expanded up to 
the relative order of $(Z\alpha)^2$. This correction 
can be rewritten as the sum of several effective operators acting on the 
nonrelativistic wave function. Then, we have an $F''_1(0)$ correction
evaluated on the nonrelativistic wave function. From the magnetic 
form factor, we have an $F_2(0)$ correction evaluated on the 
relativistic wave function, and an $F'_2(0)$ correction 
on its nonrelativistic counterpart. The form factors are
known in both photon mass~\cite{Re1972a,Re1972b} as well as
dimensional~\cite{BoMaRe2003,BoMaRe2004} regularization.
For dimensional regularization, all the relevant formulas 
are explicitly given in \cite{JeCzPa2005}, and the 
terms corresponding to the above mentioned form factors 
are indicated. In order to go over to photon energy regularization,
one has to convert the photon mass to a noncovariant cutoff.
For the lowest-order form factor slope contributing to the 
leading $\alpha \, (Z\alpha)^4 $ correction to the Lamb shift
(in units of the electron mass), this is described in the
textbook of Itzykson and Zuber~\cite{ItZu1980}. For a general 
hydrogenic state, we use a different ansatz, 
namely a generalization of the approach 
described previously for $P$ and $D$ states in 
Refs.~\cite{JePa1996,JeEtAl2003}, in order to express the 
high-energy part as a function of $\epsilon$ for a general
state of the hydrogen atom.

There is a further two-vertex operator which is given by the 
diagrams in Fig.~5 of Ref.~\cite{JeCzPa2005}.
It corresponds to the following Hamiltonians in 
the three regularizations,
\begin{subequations}
\begin{align}
H(\epsilon) =& \; \frac{\alpha}{\pi} \, 
\left[ \frac23 \ln\left( \frac{1}{2 \epsilon} \right) -
\frac{2}{3 \epsilon} + \frac{34}{45} \right] \,
(\vec{\nabla} V)^2 \,, 
\\[2ex]
H(\mu) =& \; \frac{\alpha}{\pi} \, 
\left[ \frac23 \, \ln\left( \frac{1}{\mu} \right) 
  - \frac{3 \pi}{16 \mu} - \frac16 \right]\,
(\vec{\nabla} V)^2 \,, 
\\[2ex]
H(\varepsilon_D) =& \; \frac{\alpha}{\pi} \, 
\left( \frac16 - \frac{1}{3 \, \varepsilon_D} \right) 
(\vec{\nabla} V)^2 \, .
\end{align}
\end{subequations}

The above formulas, however, are of little use for a comparison
to experiments unless complemented by their evaluation 
on a general hydrogenic state in terms of actual quantum 
numbers. A general result for the high-energy part in dimensional 
regularization, valid for all states with nonvanishing 
angular momentum and for the weighted difference of $nS$ states
(where $n$ is the principal quantum number), has been given 
in Ref.~\cite{JeCzPa2005}. We here refer to Eq.~(3.35) {\em ibid.}, with 
partial results given in Eqs.~(3.32) and (3.34) {\em ibid.}, 
and the latter term corresponds to our $H(\varepsilon_D)$. 
This result is expressed in terms of matrix elements to be evaluated 
on the reference state, which is manifestly taken as 
a nonrelativistic Schr\"{o}dinger eigenstate. These matrix elements,
as given in~\cite{JeCzPa2005}, constitute rather complicated 
expressions and are not evaluated in terms of quantum numbers.
In photon 
energy regularization, the general form of the result for the 
high-energy part has been indicated in Eq.~(8) of Ref.~\cite{JeEtAl2003},
but the quantities $\mathcal K$ and $\mathcal C$ in that equation were
given in general form only for selected submanifolds of states.

In this Letter, we are in the position to note that 
the final results for the high-energy part, 
in all three regularizations, can be expressed in a 
very compact form for all states with orbital angular 
momentum $L \geq 2$,
\begin{subequations}
\label{FH}
\begin{align}
\label{FHeps}
F_H(\epsilon) =& \; 
\Xi + (Z\alpha)^2 
A_{61} \left[ \ln\left( \frac{1}{2 \epsilon} \right) -
\frac{1}{\epsilon} + \frac{17}{15} \right]\,, \\[2ex]
\label{FHmu}
F_H(\mu) =& \;
\Xi + (Z\alpha)^2 
A_{61} \left[ \ln\left( \frac{1}{\mu} \right) -
\frac{9}{32 \mu} - \frac14 \right]\,, 
\\[2ex]
\label{FHdim}
F_H(\varepsilon_D) =& \;
\Xi + (Z\alpha)^2 
A_{61} \left( \frac14 - \frac{1}{2 \varepsilon_D} \right) \,.
\end{align}
\end{subequations}
The $A_{61}$ coefficient is defined in Eq.~(\ref{ACES})
and can be given as ($L \geq 2$)
\begin{align}
A_{61} =& \; \frac23 \, \frac{n^3}{(Z\alpha)^4} \, 
\left< \phi \left| \frac{1}{r^4} \right| \phi \right>
\nonumber\\[1ex]
=& \; 
\frac{3 n^2 - L(L+1)}{3 n^2 \, (L+\threehalf) (L+1) (L + \half) 
L (L - \half)} \,,
\end{align}
where $|\phi \rangle$ is the Schr\"{o}dinger eigenstate.
Note that $A_{61}$ is independent of $j$ for $L \geq 2$.
The matrix element $\Xi$ is derived from the magnetic 
form factor correction to the Lamb shift and can be 
expressed either as a sum of various effective operators 
acting on the nonrelativistic hydrogenic wave function,
or as a single operator acting on the full 
relativistic Dirac wave function, appropriately expanded in 
powers of $Z\alpha$. The latter approach leads to the 
most compact expression, and the resulting matrix 
element can be related to the integral denoted as 
$C^{-2}_{n\kappa,n\kappa}$ on p.~4483 of~\cite{Sh1991} 
and evaluated using generalized virial relations 
for the Dirac equation.
Indeed, the result reads, expanded in subleading order in 
the $Z\alpha$-expansion ($\vec{E} = - \vec{\nabla} V$ is the 
electric field generated by the atomic nucleus with 
$V = -Z\alpha/r$),
\begin{align}
\label{Xi}
& \Xi = \frac{n^3}{(Z\alpha)^4} \, 
\left< \psi^+ \left| \frac{\rm i}{4} \, \vec{\gamma}\cdot\vec{E} \,
\right| \psi \right>
= - \frac{1}{2 \kappa \,(2 L + 1)} 
\nonumber\\[2ex]
&  + (Z\alpha)^2 \, \left( 
- \frac{12\kappa^2 - 1}%
{2\,(2 j + 1)\,\kappa^2\,(2 \kappa-1)\,(2 \kappa+1)^2} \right.
\nonumber\\[2ex]
& \left.
- \frac{1}{n} \, \frac{3}{4 \kappa^2 \, (2\kappa + 1)}
+ \frac{1}{n^2} \, \frac{8 \kappa - 3}
  {2 \, (2 j + 1) \, (2\kappa - 1) \, (2\kappa + 1)} \right)
\nonumber\\[2ex]
& = - \frac{1}{2 \kappa \,(2 L + 1)} + (Z\alpha)^2 \, \, \Xi_2 \,,
\end{align}
where $\psi$ is the relativistic Dirac wave function,
and $\psi^+$, for clarity, is its adjoint (row vector in 
spinor space, complex conjugated), which is different 
from the Dirac adjoint $\bar{\psi} = \psi^+ \gamma^0$.
The Dirac quantum number is $\kappa = 2 (L-j) (j+ \half)$.
We here define $\Xi_2$ to be the coefficient of the 
$(Z\alpha)^2$ term (this convention will be useful later). 
This completes our treatment of high-energy photons.

{\em {Low-energy part.---}}In a certain sense, the
photon energy regularization constitutes 
the most natural procedure for low-energy photons.
One simply expands the transition current via
a Foldy--Wouthuysen transformation~\cite{JePa1996}, and then
one applies time-independent perturbation theory 
from the low-energy terms in the resulting NRQED Hamiltonian. 
One then integrates the photon energy to some 
upper cutoff $\epsilon$ (in~\cite{Je2005mpla}, it is explained why 
the expansion first in $\alpha$, then in $\epsilon$ is
actually an expansion for large $\epsilon$).

We now describe briefly how to convert the result obtained in 
photon energy regularization to photon mass
regularization. For the leading-order term of
order $\alpha (Z\alpha)^4$, the by now famous 
substitution~\cite{Fe1949,FrWe1949,ItZu1980} reads
$\ln(\mu) \to \ln(2 \epsilon) + \textstyle{\frac56}$
while for the higher-order terms,
one has to be very careful in distinguishing
$k = |\vec{k}|$ from $\omega = \sqrt{ \vec{k}^2 + \mu^2}$.
The so-called quadrupole term obtained 
by expanding the exponential $\exp({\rm i} \vec{k}\cdot\vec{r})$
in the nonrelativistic transition current 
$p^i \, \exp({\rm i} \vec{k}\cdot\vec{r})$ is very sensitive
to the changes in the matching of $\mu$ and $\epsilon$ because
the power of the photon momentum $k$ is 
different from the nonrelativistic dipole term.
The additional terms can, however, be written 
in closed analytic form.

Finally, the full evaluation of the low-energy part in
dimensional regularization is described in detail in 
Ref.~\cite{JeCzPa2005}, and we are now in the position
to indicate the results as follows.
We denote the (nonrelativistic) Bethe logarithm 
by $\ln k_0$ and the 
relativistic Bethe logarithm by ${\mathcal L}$,
following the conventions of Ref.~\cite{JeEtAl2003,JeCzPa2005}.
Both of these quantities are of course state dependent,
and they can both be evaluated only numerically.
In the three different regularizations, the results read 
(for states with angular momenta $L \geq 2$)
\begin{subequations}
\label{FL}
\begin{align}
\label{FLeps}
& F_L(\epsilon) = \; - \frac43 \ln k_0
+ (Z\alpha)^2 \, \biggl\{ A_{61} \,
\left[ 
\ln\left( \frac{\epsilon}{(Z\alpha)^2} \right) + \frac{1}{\epsilon} \right] 
+ {\mathcal L} \biggr\} \,,
\\[2ex]
\label{FLmu}
& F_L(\mu) = \; - \frac43 \ln k_0
+ (Z\alpha)^2 \, \biggl\{ A_{61} \,
\left[ \frac{83}{60} 
+ \frac{9}{32 \mu} +
\ln\left( \frac12 \, \frac{\mu}{(Z\alpha)^2} \right) \right] + {\mathcal L}
\biggr\} \,,
\\[2ex]
\label{FLdim}
& F_L(\varepsilon_D) = \; - \frac43 \ln k_0
+ (Z\alpha)^2 \, \biggl\{ A_{61} \,
\left[ \frac{53}{60} 
+ \frac{1}{2 \varepsilon_D} +
\ln\left( \half \, (Z\alpha)^{-2} \right) \right] + {\mathcal L}
\biggr\} \,.
\end{align}
\end{subequations}

\newcommand{\makespace}{ \rule[-3.5mm]{0mm}{9mm} }

\begin{table*}[htb!]
\begin{center}
\begin{scriptsize}
\caption{\label{table1} Explicit high- and low-energy parts
for the $8D_{3/2}$ state. The $F_H$ is the 
contribution to the self-energy correction from the high-energy
part, the $F_L$ is the low-energy part, and the three
regularizations are: $\epsilon$ denotes the photon energy,
$\mu$ denotes the photon mass, and in dimensional 
regularization, we work in $4 - 2 \varepsilon_D$ space-time 
dimensions.}
\begin{tabular}{c}
\hline
\hline
\makespace
$\displaystyle{ F_{\rm H}(8{\rm D}_{3/2}, \epsilon) =
- \frac{1}{20} +
(Z\alpha)^2 \, \left[-\frac{20893}{2419200}
- \frac{31}{2520\,\epsilon}
- \frac{31}{2520} \, \ln(2\epsilon) \right] }$ \\
\makespace
$\displaystyle{ F_{\rm L}(8{\rm D}_{3/2}, \epsilon)
= - \frac43\,\ln k_0 (8 {\rm D}) +
(Z\alpha)^2 \, \left[0.024\,886
+ \frac{31}{2520\,\epsilon}
+ \frac{31}{2520} \, \ln\left(\frac{\epsilon}{(Z\alpha)^2}\right)
\right]}$ \\
\hline
\makespace
$\displaystyle{ F_{\rm H}(8{\rm D}_{3/2}, \mu) =
- \frac{1}{20} +
(Z\alpha)^2 \, \left[-\frac{20687}{806400}
- \frac{31 \pi}{8960 \,\mu}
- \frac{31}{2520}\,\ln\left(\frac{1}{\mu}\right) \right] } $ \\
\makespace
$\displaystyle{ F_{\rm L}(8{\rm D}_{3/2}, \mu)
= -\frac43\,\ln k_0 (8 {\rm D}) +
(Z\alpha)^2 \, \left[0.033\,376
+ \frac{31 \pi}{8960 \, \mu}
+ \frac{31}{2520} \, \ln\left[\frac{\mu}{(Z\alpha)^2}\right]
\right] } $ \\
\hline
\makespace
$\displaystyle{ F_{\rm H}(8{\rm D}_{3/2}, \varepsilon_D ) =
- \frac{1}{20} +
(Z\alpha)^2 \, \left[-\frac{15727}{806400}
- \frac{31}{5040\,\varepsilon_D} \right]} $ \\
\makespace
$\displaystyle{ F_{\rm L}(8{\rm D}_{3/2}, \varepsilon_D )
= -\frac43\,\ln k_0 (8 {\rm D}) +
(Z\alpha)^2 \, \left[0.027\,226
+ \frac{31}{5040\,\varepsilon_D}
+ \frac{31}{2520} \, \ln\left[(Z\alpha)^{-2}\right]
\right] }$ \\
\hline
\makespace
(Sum $F = F_H + F_L$) 
\quad $\displaystyle{ F(8{\rm D}_{3/2}) =
- \frac{1}{20} - \frac43\,\ln k_0 (8 {\rm D})
+ (Z\alpha)^2 \, \left[
\frac{31}{2520} \, \ln\left[(Z\alpha)^{-2}\right]
+ 0.007\,723 \right]}$ \\
\hline
\hline
\end{tabular}
\end{scriptsize}
\end{center}
\end{table*}

{\em Adding the high- and low-energy parts.---}It 
is easy to see that when adding the 
high- and low-energy parts from Eqs.~(\ref{FH}) and~(\ref{FL}),
not only the regularization parameters cancel, 
but also, a reparameterization-invariant result is 
obtained,
\begin{equation}\label{F}
\begin{split}
F &= - \frac{1}{2 \kappa (2 L + 1)} - \frac43 \, \ln k_0
\\
& \quad + (Z\alpha)^2 \, \left\{ A_{61} \, 
\left[ \ln\left( \half (Z\alpha)^{-2} \right) + \frac{17}{15} \right] + 
\Xi_2 + {\mathcal L} \right\} \,.
\end{split}
\end{equation}
The reparameterization invariance of NRQED
is thus verified in a nontrivial 
calculation beyond leading order,
in all three common regularization methods.
A concrete numerical example is given in Table~\ref{table1},
where the explicit numerical coefficients are written
out for the $8{\rm D}_{3/2}$ state (this hydrogenic level
is spectroscopically important~\cite{BeEtAl1997}).

Having obtained compact expressions, the question 
can be asked whether it is possible to 
evaluate, beyond leading order, the relativistic 
Bethe logarithms ${\mathcal L}$ for highly excited states of 
hydrogenlike atoms, in approximately the same way as
for the nonrelativistic counterparts
(the ``usual'' Bethe logarithms), for which a 
systematic investigation has been started in 
Ref.~\cite{DrSw1990} in relation to excited states.
In order to appreciate the difficulties associated with the problem, 
one should recall that the relativistic Bethe logarithms 
represent a comparatively much more demanding 
calculation as compared to their nonrelativistic counterparts,
and the first such evaluation was not done until 1993 
(see Ref.~\cite{Pa1993}), i.e.~46 years after the 
evaluation of the nonrelativistic counterpart~\cite{Be1947}. 

Analytic and semi-analytic calculations,
where all expressions are kept in full analytic 
form before the final photon energy integration,
are prohibitively difficult for states with higher principal
quantum numbers, as already described in a number of previous 
works on the subject of interest. It is doubtful if the 
analytic approach to the evaluation of matrix elements
with the hydrogenic propagator, which is commonly based on a
Sturmian decomposition~\cite{SwDr1991a,SwDr1991b,SwDr1991c},
can ever be generalized beyond principal quantum 
number $n = 8$, where on the order of $10^5$ terms are
encountered in intermediate steps~\cite{JeEtAl2003}. 
Calculations for the relativistic corrections
to higher excited states seem to be possible only via
completely numerical (lattice) methods.

\begin{table}[htb!]
\begin{center}
\begin{minipage}{6.5cm}
\begin{center}
\caption{\label{table2} 
Relativistic Bethe Logarithms 
${\mathcal L}$ and $A_{60}$ 
coefficients for highly excited $D$ states.}
\newcolumntype{e}{D{.}{.}{11}}
\begin{tabular}{r@{\hspace{0.2in}}e@{\hspace{0.2in}}e}
\hline
\hline
\rule[-2mm]{0mm}{6mm}
$n$ & 
\multicolumn{1}{l}{${\mathcal L}(nD_{3/2})$} & 
\multicolumn{1}{l}{${\mathcal L}(nD_{5/2})$} \\
\hline
 9 & 0.025~043~91(5) & 0.022~564~66(5) \\
10 & 0.025~185~92(5) & 0.022~669~65(5) \\
11 & 0.025~280~93(5) & 0.022~733~86(5) \\
12 & 0.025~353~59(5) & 0.022~780~80(5) \\
\hline
\hline
\rule[-2mm]{0mm}{6mm}
$n$ &
\multicolumn{1}{l}{$A_{60}(nD_{3/2})$} &
\multicolumn{1}{l}{$A_{60}(nD_{5/2})$} \\
\hline
 9 & 0.008~083~01(5) & 0.034~735~88(5) \\
10 & 0.008~413~79(5) & 0.034~832~71(5) \\
11 & 0.008~681~09(5) & 0.034~876~38(5) \\
12 & 0.008~909~60(5) & 0.034~896~67(5) \\
\hline
\hline
\end{tabular}
\end{center}
\end{minipage}
\end{center}
\end{table}

Here, a numerical approach inspired by 
a discretized space as used by Salomonson and Oester~\cite{SaOe1989}
is used, and up to eleven-point discretized representations are used 
in order to represent differential operators 
on the lattice whose coordinates are chosen to 
represent very accurately the origin in coordinate space. 
Values for the relativistic Bethe logarithms 
${\mathcal L}$ and for the $A_{60}$ coefficients of highly excited D 
states are given in Table~\ref{table2},
where we note that the $12{\rm D}_{3/2}$ and
$12{\rm D}_{5/2}$ states are of particular
experimental interest~\cite{ScEtAl1999}. 

{\em {Conclusions.---}}In summary, we have completed two goals in this
Letter. (i) The reparameterization invariance of NRQED
has been verified through relative order $(Z\alpha)^2$ for 
a rather fundamentally important QED correction to the
spectrum of hydrogenlike atoms: namely, the one-photon
self-energy for excited states in a hydrogenlike 
system with orbital angular momentum quantum number
$L \geq 2$. It has been verified that the photon energy,
the photon mass and the dimensional regularizations give the same
results for the energy shift [see Eqs.~(\ref{FH}), (\ref{FL}) 
and (\ref{F})]. Because the higher-order binding corrections
to the Lamb shift involve a multitude of terms, this 
fact is rather nontrivial and is displayed in a particularly 
clear manner in the compact expressions 
for the self-energy effects obtained here. 
(ii) Numerical techniques for the calculation
of the relativistic Bethe logarithm ${\mathcal L}$
have been developed which circumvent problems associated 
to the growth of the number
of terms in intermediate steps with the principal
quantum number; these problems otherwise prohibit
analytic and semi-analytic evaluations for highly 
excited states. With the methods described here, 
calculations become possible for 
Rydberg states of the hydrogen atom, and these
are important for ultra-high-precision
spectroscopy~\cite{BeEtAl1997,ScEtAl1999}.

The two above mentioned aspects are important 
for two rather diverse topics:
(i) for a fundamental reassurance regarding the 
internal consistency of NRQED and the 
consistency of overlapping 
parameters used in field theories in general (ii) for obtaining 
improved theoretical predictions for transition
frequencies in hydrogenlike atoms.

{\em Acknowledgments.---} The authors acknowledge
helpful conversations with Professors Krzysztof Pachucki and
Peter J. Mohr. U.D.J.~acknowledges
support from Deutsche Forschungsgemeinschaft
(Heisenberg program).

\end{document}